\RequirePackage[hyphens]{url}
\documentclass[letterpaper,twocolumn,10pt]{article}
\usepackage{template/usenix-template}

\usepackage{amssymb} 
\usepackage{pifont}  
\usepackage{xcolor}  
\usepackage{array}
\usepackage{xspace}
\usepackage{subfigure}
\usepackage{caption}
\usepackage{multirow}
\usepackage{tcolorbox}

\usepackage{hyperref}
\usepackage{textcomp}
\usepackage{enumitem}
\usepackage[normalem]{ulem}

\renewcommand{\emph}[1]{\textit{#1}}

\newcommand{\sys}{PulseNet\xspace} 
\newcommand{\wlet}{Pulselet\xspace} 

\ifdefined\RELEASE
  \newcommand{\ignore}[1]{}
  \newcommand{\fixme}[1]{}
  \newcommand{\dmi}[1]{}
  \newcommand{\leo}[1]{}
  \newcommand{\hc}[1]{}
  \newcommand{\lazar}[1]{}
  \newcommand{\bx}[1]{}
  \newcommand{\dha}[1]{}
  \newcommand{\kev}[1]{}
  \newcommand{\hon}[1]{}
  \newcommand{\TODO}[1]{}

\else
  \newcommand{\ignore}[1]{}
  \newcommand{\fixme}[1]{{\textcolor{red}{[~FIXME:~#1~]}}}
  \newcommand{\dmi}[1]{{\textcolor{blue}{[~D:~#1~]}}}
  \newcommand{\leo}[1]{{\textcolor{teal}{[~Le:~#1~]}}}
  \newcommand{\hc}[1]{{\textcolor{gray}{[~H:~#1~]}}}
  \newcommand{\lazar}[1]{{\textcolor{olive}{[~LA:~#1~]}}}
  \newcommand{\bx}[1]{{\textcolor{brown}{[~B:~#1~]}}}
  \newcommand{\dha}[1]{{\textcolor{orange}{[~Dh:~#1~]}}}
  \newcommand{\kev}[1]{{\textcolor{orange}{[~Ke:~#1~]}}}
  \newcommand{\hon}[1]{{\textcolor{orange}{[~Ho:~#1~]}}}

  \newcommand{\TODO}[1]{{\textcolor{red}{TODO:~#1}}}
  
\fi

\begin{document}

\title{Melding the Serverless Control Plane with the Conventional Cluster Manager for Speed and Resource Efficiency}

\author{
{\rm Leonid Kondrashov}\\
NTU Singapore
\and
{\rm Lazar Cvetkovi\'{c}}\\
ETH Zurich
\and
{\rm Hancheng Wang}\\
Nanjing University
\and
{\rm Boxi Zhou}\\
NTU Singapore
\and
{\rm Dmitrii Ustiugov}\\
NTU Singapore
}


\maketitle


\begin{abstract}

Serverless platforms face a trade-off: conventional cluster managers like Kubernetes offer compatibility for co-locating Function-as-a-Service (FaaS) and Backend-as-a-Service (BaaS) components of serverless applications, at the cost of high cold-start latency, whereas specialized FaaS-only systems like Dirigent achieve low latency by sacrificing compatibility, preventing integrated management and optimization. Our analysis reveals that FaaS traffic is bimodal: predictable, sustainable traffic consumes >98\% of cluster resources, whereas sporadic, excessive bursts stress the control plane's scaling latency, not its throughput.

With these insights, we design \emph{\sys}, a serverless architecture that uses a dual-track control plane tailored to both traffic types. \sys's standard track manages sustainable traffic with long-lived, full-featured Regular Instances under a conventional cluster manager, preserving compatibility for the majority of the workload. To handle excessive traffic, an expedited track bypasses the slow manager to rapidly create short-lived, disposable Emergency Instances, minimizing cold-start latency and resource waste from idle instances. This hybrid approach achieves 35\% better performance than Dirigent, a FaaS-only system, on a production workload at the same cost and outperforms other Kubernetes-compatible systems by 1.5–3.5$\times$, reducing the cost by up to 70\%.

\end{abstract}

\section{Introduction}
\label{sec:intro}

Serverless cloud applications today comprise stateless Function-as-a-Service~(FaaS) and conventional stateful Backend-as-a-Service~(BaaS) components, organized as DAG workflows. Developers deploy their business logic as functions while offloading the management of the underlying cloud infrastructure, for both FaaS and BaaS, entirely to the providers. 
The control plane is key to providing fast and timely scaling
and resource efficiency when running these applications at the massive cloud scale, which can be achieved by careful placement and load balancing, and timely scaling.

Unfortunately, despite the tight coupling of FaaS and BaaS components in cloud applications, their control planes lack coordination, often rendering co-design and optimization impractical. For example, AWS Lambda~\cite{agache:firecracker} and Dirigent~\cite{cvetkovic:dirigent} manage their own fleets of bare-metal nodes allocated exclusively for FaaS. Other works focus on designs for specific BaaS services~\cite{klimovic:pocket,mahgoub:sonic} used by serverless applications, such as storage and database.
In contrast, a decade ago, the success of Google's Borg cluster manager demonstrated the benefits of running various services under a single control plane, achieving better cluster utilization and reducing operational costs without compromising performance~\cite{verma:borg,tirmazi:borg}. 
Since then, practitioners have designed many open-source production systems~\cite{openfaas,knative,fission,nuclio} 
running atop Borg's successor, Kubernetes~\cite{k8s}, which features numerous optimizations and highly-optimized technologies (e.g., intelligent placement, load balancing, networking, and authentication) and can colocate any services in the same cluster.
However, state-of-the-art prior works~\cite{cvetkovic:dirigent,roy:icebreaker,singhvi:atoll} have concluded that a heavyweight control plane, as in Kubernetes, is too slow to keep up with highly dynamic FaaS workloads~\cite{shahrad:serverless}, calling for a full revamp of technologies already available for BaaS services running under a conventional control plane.

In contrast, we show that a conventional cluster manager's control plane, although heavyweight, can be seamlessly extended to provide sufficient scaling speed for FaaS workloads, thereby unlocking efficient co-location of FaaS and BaaS systems without compromising performance.
Using production Azure Functions traces~\cite{shahrad:serverless}, we present a thorough analysis of the FaaS workload and performance characteristics in systems representative of leading commercial offerings~\cite{gcr,aws-lambda-scaling} and open-source systems~\cite{openwhisk,knative_offerings}.
First, we identify two kinds of invocation traffic,
\emph{sustainable} and \emph{excessive}, showing that the former utilizes $>$98\% of the cluster CPU resources while the latter stresses the control plane but consumes $<$2\% of the cluster resources. 
Second, we demonstrate that control planes suffer from long instance-creation delays due to frequent interaction with the centralized cluster manager, e.g., when allocating cluster resources and IP addresses, and setting up network routes. However, we show that these systems' control planes can be tuned to deliver throughput sufficient for a dynamic large-scale FaaS deployment, in contrast to the prior work~\cite{cvetkovic:dirigent} that has optimized for control-plane performance. 
Finally, we identify that current systems waste 70-87\% of memory and 9-20\% of CPU resources due to prolonged idle instance lifetime and high instance churn, respectively.

Based on the obtained insights, we introduce \emph{\sys},\footnote{We will release the code, traces, and toolchain after publication.} a serverless system designed to achieve high performance and low cost while maintaining compatibility with conventional cluster managers, such as Kubernetes~\cite{k8s}. 
\emph{\sys} employs a novel dual-track control plane comprising \emph{standard} and \emph{expedited} tracks. 
The standard track is based on Knative atop Kubernetes, the industry-standard serverless cluster manager,
managing long-lived, full-featured \emph{Regular Instances} to handle the sustainable traffic. This track adjusts the number of instances off the critical path in cooperation with the underlying cluster manager, thereby preserving full compatibility with its rich feature set. Concurrently, the expedited track serves the excessive traffic bursts by rapidly creating \emph{Emergency Instances} that completely bypass the cluster manager and its bookkeeping.
This track communicates directly with a node-local agent that spawns single-use instances, which it shuts down after processing a single invocation, minimizing idle instance lifetime and associated resource waste. These disposable instances are fully compatible with FaaS workloads, albeit they forego the conventional cluster manager's features for long-running instances, such as readiness and liveliness probes, and advanced networking capabilities, which they do not need, in exchange for much faster instance creation.


\newcommand{\cmark}{\textcolor{green}{\checkmark}} 
\newcommand{\xmark}{\textcolor{red}{\ding{55}}}  
\newcolumntype{C}[1]{>{\centering\arraybackslash}m{#1}}

\begin{table}
\centering

\centering
\begin{tabular}{| p{0.34\linewidth} | C{0.08\linewidth} | C{0.06\linewidth} | C{0.11\linewidth} | C{0.08\linewidth} | C{0.1\linewidth} | C{0.13\linewidth} |}
\hline
System & React. time & CM Perf. & Predict. Comp. & FaaS-BaaS coloc. & Resour. Waste\\
\hline
\hline
AWS Lambda (sync) & \cmark & \cmark & \xmark & \xmark & High \\
\hline
OpenWhisk (sync) & \cmark & \xmark & \xmark & \cmark & High \\
\hline
Knative (async) & \xmark & \xmark & \cmark & \cmark & Moder. \\
\hline
Dirigent~\cite{cvetkovic:dirigent} & \xmark & \cmark & \cmark & \xmark & Low \\
\hline
\hline
\sys~(ours) & \cmark & \cmark & \cmark  & \cmark & Low \\
\hline

\end{tabular}
\vspace{-10pt}
\caption{
Comparison of the existing approaches and the proposed \sys in reaction time (React. time), cluster manager performance (CM Perf.), compatibility with predicting models (Predict. Comp.), ability to colocate FaaS and BaaS (FaaS-BaaS coloc.), and Resource Waste.}
\vspace{-18pt}
\label{table:system_comparison_table}
\end{table}

\begin{figure}
    \centering
    \includegraphics[width=0.9\linewidth]{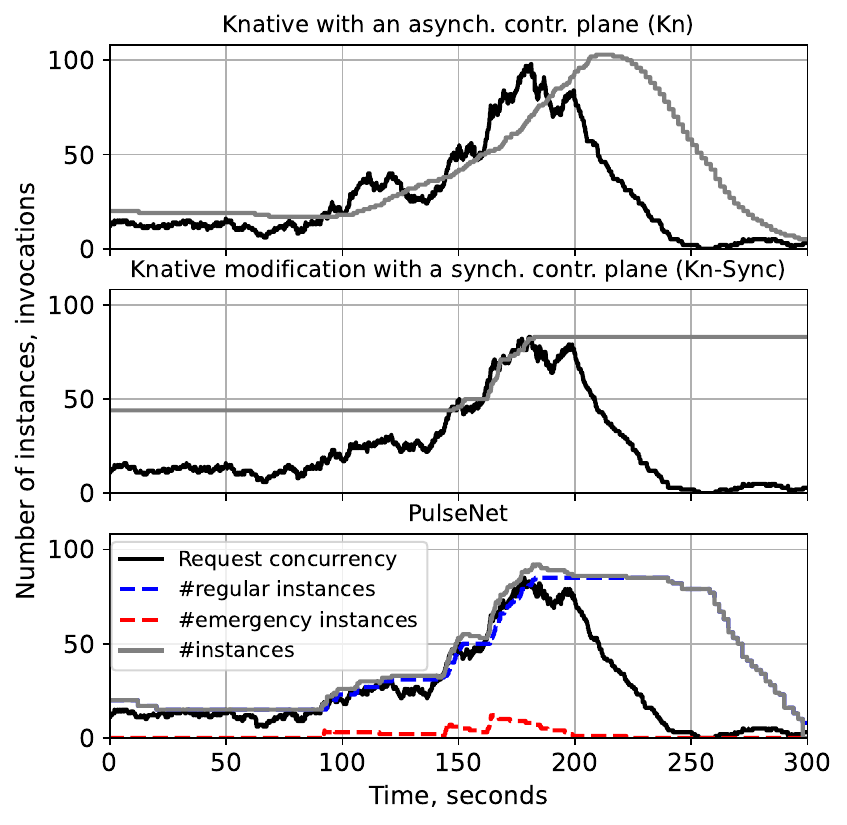}  
    \vspace{-10pt}
    \caption{Instance number scaling over time in response to the changes in the in-flight request concurrency in the state-of-the-art systems and \sys.
    Kn scales too slowly, whereas Kn-Sync incurs high costs by keeping instances idle for prolonged periods. Dirigent's behavior is similar to Kn (not shown). 
    }
    \vspace{-15pt}
    \label{fig:timelines}
\end{figure}


We prototype \sys in vHive~\cite{ustiugov:benchmarking} as a seamless extension of the Knative control plane running atop unchanged Kubernetes. 
Evaluation with a production workload shows that \emph{\sys} provides 35\% performance improvement while maintaining cost parity with Dirigent~\cite{cvetkovic:dirigent}. 
Additionally, when compared with other Kubernetes-compatible alternatives, \emph{\sys} delivers 1.5-3.5$\times$ better performance than the systems with a synchronous control plane, as in AWS Lambda~\cite{ray-serve-autoscaling}, while reducing the operational cost by 8-70\%; and achieves 1.7-3.5$\times$ better performance than the systems with an asynchronous control plane, as in Knative, while reducing the cost by 3-65\%. Finally, compared with the systems featuring a linear regression and the state-of-the-art NHITS prediction models~\cite{challu:nhits,joosen:how}, \emph{\sys} delivers up to 4$\times$ better performance while reducing costs by 35-40\%. 

We summarize the comparisons across state-of-the-art systems in Table~\ref{table:system_comparison_table} and illustrate the scaling speed and resource efficiency in Figure~\ref{fig:timelines} on a trace snippet taken from Azure Functions trace~\cite{shahrad:serverless}. In contrast to the baselines, \sys closely tracks traffic changes by rapidly creating emergency instances whenever the standard track lags behind and quickly tearing them down once the traffic trend settles.




Our main contributions are: 
\vspace{-8pt}
\begin{itemize}[leftmargin=0cm,itemindent=.2cm,labelwidth=\itemindent,labelsep=0cm,align=left]
\setlength\itemsep{0pt}
    \item We identify two types of invocation traffic, \emph{sustainable} and \emph{excessive}, with the former consuming $>$98\% of the cluster resources and the latter stressing the control plane.
    
    \item This work is the first to \emph{comprehensively} characterize the performance and cost in systems with synchronous and asynchronous control planes.
    We also show that previously proposed intelligent scaling predictors~\cite{challu:nhits,joosen:how} incur a substantial cost overhead, with nearly 5$\times$ resource overprovisioning.
    
    \item We propose \sys, a novel \emph{dual-track} control-plane architecture that unlocks fast, resource-efficient scaling of FaaS workloads while retaining full compatibility with conventional cluster managers and their rich feature set.
    
    \item \sys outperforms Kubernetes-compatible \emph{and} FaaS-specialized systems~\cite{cvetkovic:dirigent} by 1.5-3.5$\times$ with a sampled production workload, reducing cost by up to 70\%, when running synthetic and real-world benchmarks. 

    
\end{itemize}

\section{Background}
\label{sec:back}

Here, we describe two serverless application deployment approaches, disaggregated and co-located, and detail the state-of-the-art control-plane architectures and associated trade-offs. 



\subsection{FaaS\&BaaS: Disaggregated or Co-Located}
\label{sec:back_disag_vs_coloc}

Serverless applications comprise functions~(FaaS) along with conventional microservices~(BaaS), e.g., for cross-function communication.
To date, cloud providers have followed two deployment approaches: disaggregated with FaaS and BaaS components running in separate clusters with independent FaaS-~\cite{cvetkovic:dirigent,singhvi:atoll,mvondo:ofc,fuerst:faascache} and BaaS-specialized~\cite{romero:faast,klimovic:pocket,mahgoub:sonic} control planes, vs. co-located with both FaaS and BaaS orchestrated by the same cluster manager~\cite{fission, knative, openfaas, openwhisk}.

Following the disaggregated approach, system architects design the cluster manager tailored for highly sporadic FaaS workloads, unlocking orders of magnitude faster scaling speed, which conventional, microservice-centric cluster managers cannot deliver~\cite{cvetkovic:dirigent}. 
However, despite the promised scaling-speed benefits, foregoing compatibility with the conventional cluster manager requires the cumbersome re-implementation of many essential features that academic prototypes often overlook, e.g., authentication, DNS, and service mesh deployments for West-East traffic~\cite{service_mesh}, and rules out affinity-based optimizations~\cite{kraft:apiary, s3objectlambda}. 



In contrast, the colocation approach is fundamentally superior, as it jointly manages FaaS and BaaS components, significantly reducing overheads and cost.
Two notable examples that showcase colocation benefits are Nightcore~\cite{jia:nightcore}, which exploits shared memory for low-latency communication, 2$\times$ faster than with RPC, 
and SPRIGHT~\cite{qi:spright}, which reduces the data plane's CPU overhead with eBPF by an order of magnitude. These works rely on locality-aware scheduling, which is hard to implement across clusters.
Besides, many open-source production systems adopt the colocation approach~\cite{fission, knative, openfaas, openwhisk}, as it allows them to benefit from Kubernetes' integrated management and placement optimizations~\cite{zhang:faster,bazzaz:preventing,verma:borg}, support for monitoring~\cite{prometheus,grafana,opentelemetry,fluentd}, network control and mounts~\cite{calico,k8s_pv}, and versioned deployments~\cite{k8s_revisions}.

The key obstacle to unlocking same-cluster management is designing an efficient joint control plane that meets the scaling-speed requirements of FaaS workloads while retaining the rich optimization and feature space of conventional managers like Kubernetes -- which is the focus of our work.

\subsection{FaaS Control-Plane Architectures}
\label{sec:back_faas_cp}


\begin{figure}
    \centering
    \subfigure[System with a synchronous control plane.]{\includegraphics[width=0.9\linewidth]{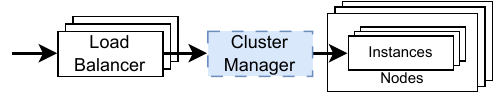}\label{fig:autoscaling:cold_sync}} 
    \subfigure[System with an asynchronous control plane.]{\includegraphics[width=0.9\linewidth]{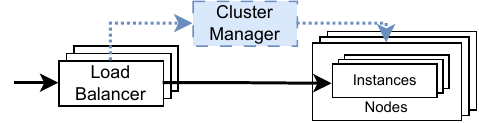}\label{fig:autoscaling:cold_async}} 
    \vspace{-10pt}
    \caption{High-level overview of serverless architecture. The cluster manager is on the path only for cold invocations.
    }
    \vspace{-18pt}
    \label{fig:cm_design}
\end{figure}

To understand the bottlenecks in current FaaS control planes, we detail the control-plane architectures and operations, similar to those used by leading commercial and open-source systems.
Figure~\ref{fig:cm_design} overviews two of the most common serverless system architectures, both of which comprise a load balancer, a cluster manager, and worker nodes that run function instances. Load balancers and function instances constitute the data plane. Serverless function invocations arrive from the application users to the scaled-out load balancers, which route invocations to function instances for execution. 
If a function has no available instances to process the invocation, the cluster manager creates more instances.

Control-plane architectures 
in state-of-the-art commercial~\cite{awslambda,knative_offerings} and open-source systems~\cite{knative,openfaas,openwhisk} tend to fall into two categories based on whether they create instances synchronously (on the critical path) or asynchronously (off the critical path). The synchronous approach, employed by production systems such as AWS Lambda~\cite{awslambda} and some open-source systems~\cite{openwhisk}, allows for an immediate scaling reaction as soon as an invocation arrives. There, the cluster manager issues an instance creation command on the critical path of an invocation to the worker node, binding the invocation for execution in the instance freshly started on that node.



Other systems, including Knative~\cite{knative}, Google Cloud Run~\cite{gcr}, and FunctionGraph~\cite{liu:gap}, use the asynchronous approach associated with a slower reaction time: these systems typically aggregate function invocation statistics over a period of time and only start scaling when they can confirm the change in the invocation traffic.
Hence, in the worst case, an invocation can wait for the entire autoscaling period (which is 2 seconds in Knative~\cite{knative:autoscaling_period}) before the cluster manager requests an instance creation. 
Moreover, prior works~\cite{liu:gap,cvetkovic:dirigent} have found that asynchronous approaches may exhibit higher tail latencies, e.g., related to their higher queuing delays in the control plane.
For example, when a single instance serves invocations,  
all invocations have to wait until the control plane decides and starts creating more instances. 
\section{Characterization of the Traffic Patterns \& Existing Control Planes' Performance}
\label{sec:char}


In this section, we 
first study the traffic patterns occurring in production deployments~(\S\ref{sec:char_inv_patterns}). Then, we study the system implications of these patterns, decomposing the scaling delays in existing state-of-the-art systems with synchronous and asynchronous control planes~(\S\ref{sec:char_cp_delays}). We also evaluate the control planes' instance-creation throughput~(\S\ref{sec:char_thru}) and cluster-resource usage efficiency~(\S\ref{sec:char_res_eff}).

We use vHive~\cite{ustiugov:benchmarking} research framework to configure two setups based on Knative~\cite{knative} and Kubernetes~\cite{k8s} for our experiments, which are widely used in commercial serverless deployments~\cite{knative_offerings,cncf:survey_2020}.
The first configuration uses vanilla Knative
that features an asynchronous control plane. 
The second configuration is a modified Knative version that features a synchronous control plane, with an autoscaling policy similar to AWS Lambda~\cite{agache:firecracker,ustiugov:stellar,aws-lambda-scaling}.
Section \ref{sec:method} provides more details about these configurations and the evaluation methodology that uses sampled production traces~\cite{azure_trace,ustiugov:enabling}.

\subsection{Invocation Traffic Patterns}
\label{sec:char_inv_patterns}






To identify key traffic patterns and system requirements, we simulate 
a serverless system with a synchronous control plane, and a keep-alive period of 10 minutes when replaying an hour-long production trace~\cite{shahrad:serverless,azure_trace}.
We develop a simulator that models request concurrency (i.e., the total number of in-flight function invocations) and the number of active and idle instances at each moment in time. We assume instantaneous instance scaling and that each instance runs on one CPU core.

We observe that FaaS traffic has two distinct components. The bulk of the traffic (99.9\%), further referred to as \emph{sustainable}, is handled by alive instances without any involvement of the control plane. Sustainable traffic uses over 98\% of the CPU resources. Only the remaining 0.1\% of traffic, which we term \emph{excessive}, triggers the creation of new instances, causing most of the load in the control plane.
Our observation corroborates the data from a major cloud provider~\cite{joosen:serverless}, reporting that only 0.01\% of invocations trigger cold starts.

This dichotomy motivates for a dual optimization strategy tailored to the distinct needs of each traffic component. For sustainable traffic, which dominates cluster resource consumption, the control plane must prioritize cost efficiency, leveraging features such as careful placement and background scaling to optimize cluster utilization. Conversely, excessive traffic consists of latency-sensitive bursts that require fast scaling.
Crucially, excessive traffic accounts for a tiny fraction of total resources ($2\%$), thus the system can trade off resource efficiency for lower latency when handling these requests without bloating overall operational costs.

\subsection{Control-Plane Scaling Delays Analysis}
\label{sec:char_cp_delays}

\begin{figure}
    \centering
    \includegraphics[width=\linewidth]{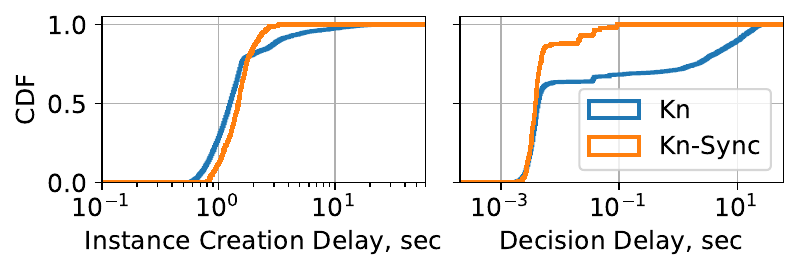}
    \vspace{-20pt}
    \caption{Cumulative distribution functions~(CDFs) for the components of delays
    occurring in the systems with synchronous (Kn-Sync) and asynchronous (Kn) control planes. 
    }
    \label{fig:scale_delays}
    \vspace{-12pt}
\end{figure}

\begin{figure}
    \centering
    \includegraphics[width=0.9\linewidth]
    {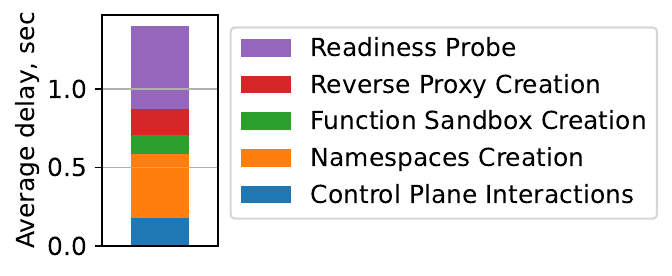}
    \vspace{-10pt}
    \caption{
    Instance creation time breakdown in Knative.
    }
    \vspace{-15pt}
    \label{fig:k8s_inst_creation}
\end{figure}




To understand the implications of the workload characteristics we discussed above, we analyze the delays that occur in the control plane of real-world synchronous and asynchronous Knative-based systems under load.
In these experiments, we use the In-Vitro~\cite{ustiugov:enabling} load generator that replays a sampled production trace containing 400 functions.


We identify and study two sources of delays in serverless control planes: the instance creation delays and the decision delays. We plot cumulative distribution functions~(CDFs) for each source in Figure~\ref{fig:scale_delays} for vanilla Knative (Kn) and its modified synchronous analogue (Kn-Sync).
The experiment shows that the component of the
queuing delays in the range from milliseconds to 10s of seconds, i.e., comparable to the invocation execution time range~\cite{shahrad:serverless}. Hence, if any of those delays happen on the critical path of invocation handling, they would be seen as noticeable delays to end-to-end request latency. 
We discuss each delay source in detail below.

\subsubsection{Instance Creation Delay}
We find that instance creation takes 1–3s to complete, a delay dominated by features in conventional cluster managers like Kubernetes (Figure~\ref{fig:k8s_inst_creation}). 
\emph{Readiness probes} introduce a 500ms average delay, a result of the 1-second minimum polling interval Kubernetes uses to protect the control plane at scale. \emph{Namespace and networking setup} requires multiple round-trips to the cluster manager, adding over 400ms. \emph{Reverse (Queue) Proxy} and function sandbox creation contribute over 250ms.
Finally, the Golang runtime in the function images we deploy has minimal initialization overhead, but other runtimes like Java can compound this delay by seconds~\cite{ustiugov:stellar}.

\subsubsection{Decision-Making Delay}

We find that decision-making latency is fast in most cases (65-85\% under 10ms), particularly when creating the first instance of a function. However, when scaling functions that are already active, confirming the traffic trend by averaging arrivals over a time window creates a long tail delay of up to 20 seconds, especially for the asynchronous control plane. We find that adjusting the number of instances, i.e., scaling not from zero, often leads to such delayed decisions, as the control plane takes time to confirm the changing trend in the traffic by averaging the arrival rate over a time window (1 minute by default).

In summary, instance creation delays are the biggest driver of median latency, prompting a redesign of the critical path. While decision-making has a low median latency, system architects should focus on optimizing its tail. An ideal system could achieve <200ms end-to-end scaling with fast instance creation, quick decision-making, and no internal congestion.

\subsection{Conventional Control-Plane Throughput}
\label{sec:char_thru}


\begin{figure}
    \centering
    \includegraphics[width=0.85\linewidth]{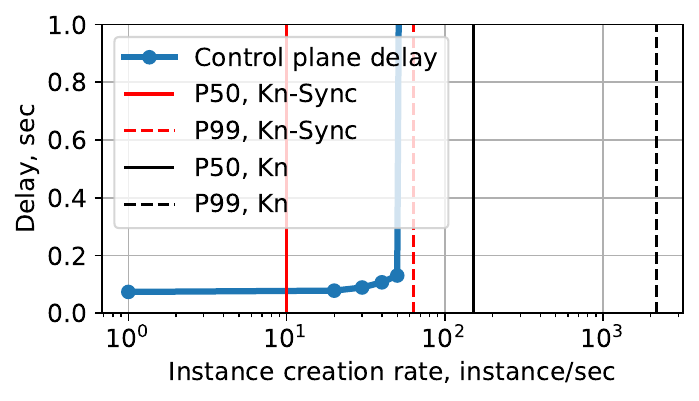}
    \vspace{-10pt}
    \caption{
    The delays occurring in the Knative control plane under various instance-creation rates, measured with a microbenchmark. The red and black lines show the required instance creation rates at the 50\textsuperscript{-th} and 99\textsuperscript{-th} percentiles, respectively, when replaying invocations from a sampled production trace in simulated synchronous~(Kn-Sync) and asynchronous~(Kn) control planes.
    }
    \vspace{-15pt}
    \label{fig:supported_cold_start_rate}
\end{figure}



We evaluate if a conventional control plane can handle the high instance creation rates of FaaS deployments~\cite{shahrad:serverless}. Using a microbenchmark on a tuned Knative-Kubernetes control plane with emulated worker nodes via KWOK~\cite{kwok}, we test its throughput limits (\S\ref{sec:method}).

Our tuned control plane sustains 50 cold starts/sec, a 25$\times$ increase over the throughput reported by prior work~\cite{cvetkovic:dirigent}\footnote{This is not a contradiction: our careful configuration and tuning, discussed in \S\ref{sec:method}, have substantially increased Knative control plane's throughput.} (Figure~\ref{fig:supported_cold_start_rate}). However, when compared to production trace demands, this throughput is 3$\times$ lower than the median rate required by an asynchronous system, confirming it can be a bottleneck~\cite{cvetkovic:dirigent}. While sufficient for a synchronous system's median load, rare 99\textsuperscript{-th} percentile bursts demand 1.2$\times$ to 40$\times$ more throughput, overwhelming either control plane.
\subsection{Cluster Resource Efficiency Analysis}
\label{sec:char_res_eff}


We analyze CPU and memory usage -- a major operational cost~\cite{agache:firecracker} -- in our synchronous and asynchronous Knative systems in the same setup as in (\S\ref{sec:char_cp_delays}). We find high overheads from two sources: idle instances consume 70\% (sync) and 87\% (async) of total instance memory, while the control plane's management tasks, which include activities like instance placement and bootstrapping, consume 9\% (sync) and 20\% (async) of CPU cycles across the cluster nodes.

This motivates an efficient control plane that provisions instances just-in-time, but also minimizes resource waste by reducing instance churn and avoiding a large pool of idle instances with low reuse probability.

\subsection{Summary and Takeaways}
\label{sec:char_takeaways}

\begin{itemize}[leftmargin=0cm,itemindent=.2cm,labelwidth=\itemindent,labelsep=0cm,align=left]
\setlength\itemsep{0pt}
\item \textbf{Sustainable and excessive traffic.} FaaS traffic is bimodal~(\S\ref{sec:char_inv_patterns}): \textit{sustainable} traffic consumes the majority~(>98\%) of cluster resources with low control-plane load, whereas sporadic \textit{excessive} traffic strains the control plane with instance creation requests despite using few~(<2\%) resources. A well-designed system must cater to both.


\item \textbf{Conventional cluster manager is too slow for FaaS cold-start requirements.} While these managers can sustain the required instance creation \textit{rate}, their creation \textit{delay} is too high for FaaS. This necessitates bypassing the manager for speed, while retaining its rich features (e.g., consistent resource allocation, placement, networking) to ensure compatibility and enable co-location with other services~(\S\ref{sec:back_disag_vs_coloc}).

\item \textbf{Existing control planes waste cluster resources.}
Asynchronous control planes waste CPU cycles on high instance churn, while synchronous control planes waste memory on keeping many instances idle. An efficient system should limit the churn and instance lifetime when reuse is unlikely.
\end{itemize}

\section{\sys Design}
\label{sec:design}



Based on the above insights, we organize the control plane in two loosely-coupled tracks to naturally fit the bimodal nature of the serverless traffic~(\S\ref{sec:char_inv_patterns}). The proposed system combines rapid scaling and compatibility with conventional cluster managers, unlocking efficient colocation of stateless and stateful services in the same cluster.

The first, \emph{standard}, track is tailored to sustainable traffic. 
Given that sustainable traffic accounts for the majority of the cluster resource usage, the standard track needs to carefully distribute the sustainable traffic among the existing instances, which this track needs to carefully place and scale following the sustainable-traffic trends. The conventional control plane, such as Knative, is a perfect fit for this goal, with its rich feature set and strictly consistent tracking and allocation of cluster resources~\S\ref{sec:back_disag_vs_coloc}.

The second, \emph{expedited}, control-plane track should deliver rapid scaling for sporadic bursts specific to excessive traffic. The system creates disposable, single-use instances to process these bursts, accounting for a tiny fraction of cluster resources. Hence, this track can substantially improve the speed of instance creation -- without tipping the load balance across the cluster, by safely bypassing the standard track's strictly consistent bookkeeping of cluster resource usage.

\subsection{\sys Architecture Overview}
\label{sec:des_arch}


\begin{figure}
    \centering
    \includegraphics[width=0.85\linewidth]{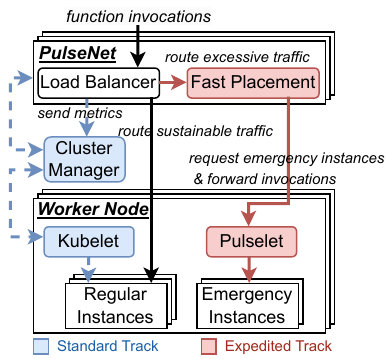}
    \caption{The \sys architecture that combines the added expedited control plane with the conventional one.
    }
    \label{fig:design}
\end{figure}


Figure~\ref{fig:design} shows \sys's workflow when processing invocations and cold starts of new function instances.
Function invocations arrive at \emph{\sys} that comprises two components: \emph{Load Balancer} and \emph{Fast Placement}.
Similar to the existing designs~(\S\ref{sec:back}), Load Balancer identifies the target function and finds a Node with an existing non-busy instance of that function to route the invocation for processing. As in systems with a vanilla asynchronous control plane, Cluster Manager continuously monitors and predicts trends in function invocation traffic, adjusting the number of instances for functions that experience up or down trends in invocation traffic. We refer to these instances as \emph{Regular Instances}. Importantly, Regular Instance creations are off the function invocations' critical path.
\sys manages Regular Instances similarly to how microservice replicas are managed in conventional managers, such as Kubernetes.

The Load Balancer chooses the track for each request based on the availability of the Regular Instances.\footnote{We consider an instance available if its per-instance queue is not full.} 
If there are available Regular Instances in the standard track, Load Balancer uses one of them. Otherwise, it sends the request to expedited track's Fast Placement component.
Fast Placement triggers an \emph{Emergency Instance} creation by sending a request to \sys's agent, called \emph{\wlet}, on one of the cluster nodes in a Round-Robin. 
\wlet spawns an Emergency Instance to process the excessive traffic. Similar to Kubernetes' kubelet~\cite{k8s} and Borg's borglet~\cite{verma:borg}, \wlet manages the lifecycle of the hosting node's Emergency Instances, shutting down each instance once it completes processing the request it was created for.

This track choice rule prioritizes optimally allocated Regular Instances, resorting to Emergency Instances in cases during traffic bursts and fast changes in the traffic trends.

\subsection{The Expedited Track: Fast, Disposable Emergency Instances}
\label{sec:des_expedited}


\sys's expedited track achieves speed by spawning Emergency Instances \emph{transparently} to the cluster manager, limiting the supported feature set, and using checkpoint-restore. \sys also lowers the cost by disposing of these instances after processing a single invocation.




\noindent\textbf{No cluster manager interaction.} \wlet does not register the Emergency Instance with the conventional cluster manager. This eliminates all associated overheads, including state persistence in etcd, scheduling, and declarative state reconciliation~\cite{cvetkovic:dirigent}.
Fast placement periodically retrieves and caches the following information from the cluster manager: function image ID, CPU and memory quotas, and revision information. Since this metadata changes only upon function redeployment or new revision roll-out.


\noindent\textbf{Limited feature set.} Emergency Instances support only the features required for processing a single function invocation:
OCI container image deployment, outbound network connections via NAT, and locally-enforced CPU and memory quotas~\cite{k8s_quotas}, foregoing the features required for long-running, reusable Regular Instances and other microservices.
First, for Emergency Instances, \sys uses local IP addresses rather than complex CNI integration for service meshes, which requires sending outbound traffic via the centralized frontend to access other services, e.g., storage.
Given that Emergency Instances account for a tiny fraction of cluster resources~(\S\ref{sec:char_inv_patterns}), this traffic cannot substantially overload the frontend.
Second, \wlet checks the readiness of Emergency Instances independently, reducing the load on the centralized cluster manager. 
Note, the features \sys supports are enough to execute arbitrary user-function code both in Regular and Emergency Instances.

\noindent\textbf{Optimized Sandbox Start.} 
\wlet uses common techniques to reduce the sandbox creation time: VM snapshot-restore and a pool of pre-initialized virtual network devices that \wlet can quickly bind to the newly restored VMs.

\noindent\textbf{Disposable Instances to Minimize Footprint}
To prevent resource waste, Emergency Instances are single-use and disposed of after they finish processing an invocation.
This ensures that the resources used to handle transient traffic bursts are reclaimed instantly, avoiding the prevalence of idle instances in the existing systems~(\S\ref{sec:char_res_eff}).
Emergency Instances process their dedicated invocations to completion and have the same scheduling priority as Regular Instances.

\subsection{The Standard Track: Efficiently Managing Sustainable Traffic with Smart Filtering}
\label{sec:des_standard}

While the expedited track ensures low-latency cold starts, the standard track is designed for resource efficiency. Its primary goal is to manage the pool of Regular Instances -- while minimizing memory waste and avoiding high instance churn that might overload the heavyweight standard track.

The key to achieving these goals is to filter the metrics coming from the expedited track, which handles the traffic spikes, before passing them to the standard track. Hence, \sys Load Balancer reports to the standard track only a fraction of invocations handled by Emergency Instances, i.e., the invocations that represent long-term trend changes rather than sporadic bursts.

In practice, we find that the function's inter-arrival time~(IAT) is a good indicator of trend consistency when compared with the keep-alive period used by the system for Regular Instances.
Thus, Load Balancer collects and periodically updates the IAT distribution for each function. 
When the Load Balancer steers an invocation to the expedited track, it also
compares the function's median IAT against \sys's keep-alive period. If the median IAT is lower than the keep-alive period,
it is worth creating an additional long-living Regular Instance for future invocations of that function. 
Hence, Load Balancer includes this invocation into the metrics reported to the cluster manager. Otherwise, the Load Balancer treats the invocation as sporadic and filters it out from the metrics.
\sys's keep-alive and filtering percentile are configurable parameters we empirically evaluate in \S\ref{sec:eval_sens}.

This filtering mechanism 
prevents the system from creating costly Regular Instances that would likely sit idle, thus reducing the overall memory usage compared to the systems that rely on long keep-alive periods~\cite{aws-lambda-scaling} or aggressive predictive scaling~\cite{joosen:how,roy:icebreaker}, as we show in \S\ref{sec:eval_memory_util}. 

\subsection{Discussion}
\label{sec:design_disc}


\noindent\textbf{Accuracy of Traffic Classification.} Despite the simplicity of our IAT-based filtering technique, it is highly efficient: \sys steers the bulk of traffic to Regular Instances~(\S\ref{sec:eval_resource_efficiency}). Misclassifications are benign: overly conservative filtering may increase the use of Emergency Instances, whereas overly permissive filtering may create more Regular Instances that are recycled by the keep-alive mechanism. We leave the design of a more intelligent traffic classifier to future work.

\noindent\textbf{Resource Allocation.} Effectively, \sys places Emergency Instances in the CPU and memory utilization margin of the cluster that is typically in the 20-60\% range in production~\cite{verma:borg,tirmazi:borg} -- more than enough to accommodate a mere 10\% fraction used by these instances in our experiments~(\S\ref{sec:eval_resource_efficiency}).

\noindent\textbf{\sys Limitations.} The \sys design relies on the observation that unpredictable bursts comprise a small fraction in production deployments, whereas the majority of function invocations can be tracked with a window-based or predictor-based autoscaler, as we show in \S\ref{sec:char_inv_patterns}, corroborating prior works~\cite{joosen:how,shahrad:serverless}. In the hypothetical scenarios where bursts dominate the traffic, \sys's expedited track might be utilized on par with the standard track, requiring a more sophisticated load balancing policy.

\section{\sys Implementation}
\label{sec:impl}




We implement \sys prototype in vHive~\cite{ustiugov:benchmarking}, an open-source framework widely used for serverless systems research in academia and industry. vHive deploys Knative~\cite{knative} production FaaS framework, used in commercial serverless offerings~\cite{knative_offerings}, atop the Kubernetes~\cite{k8s} cluster manager. We modify several Knative components, namely Activator and Autoscaler, leaving Kubernetes completely unchanged. 


\noindent\textbf{\sys Load Balancer and Metric Filtering.}
We modify the Knative Activator component to serve as a Load Balancer that routes requests to the standard track or expedited track based on the availability of Regular Instances. It also implements the metric filtering as described in~\S\ref{sec:des_standard}.

\noindent\textbf{Standard Track and Regular Instances.}
For a thorough evaluation, it is paramount to deploy a production-grade control plane for the \sys standard track to have representative overheads present in a real system. Hence, we deploy
modified Knative's control plane that follows the AWS Lambda scaling policy~\cite{aws-lambda-scaling} 
atop \emph{vanilla} Kubernetes -- with representative interactions across the control plane, cluster manager, and instances.\footnote{This approach has a limitation: Knative deploys instances in containers, rather than in MicroVMs like in commercial systems~\cite{agache:firecracker,web:gvisor}, which might show shorter sandbox startup latencies. Open-source production-grade MicroVM solutions, e.g., Firecracker or gVisor, come without a production-grade control plane, and re-implementing it in-house with a complete set of features supported by a system like AWS Lambda or Kubernetes might undermine representativity of the control-plane interactions.}

\noindent\textbf{Expedited Path: Fast Placement, \wlet, and Emergency Instances.}
To evaluate \sys dual-track design, we need to make sure that our expedited track's delays are representative of state-of-the-art systems. 
Hence, we utilize AWS Firecracker MicroVMs as sandboxes for Emergency Instances using their snapshot-restore technology, as in AWS Lambda~\cite{agache:firecracker,aws-snap-serv}.
For Fast Placement, we use a Round-Robin placement algorithm inside Knative Activator to uniformly distribute this load across all worker nodes in the cluster. 
Emergency Instances run invisibly to the cluster manager without tipping the load balance, as they account for a tiny fraction of CPU and memory resources, hence effectively running in the cluster's overprovisioning margin.
We implement \wlet in Golang and deploy it alongside the regular kubelet.

\section{Methodology}
\label{sec:method}


\noindent\textbf{Hardware setup.} 
We run \sys and other serverless systems on an 8-node c220g5 Cloudlab cluster~\cite{CloudLab}. Each node has two Intel Xeon Silver 4114 CPUs @ 2.20GHz, each with 10 physical cores, 192~GB DRAM, and an Intel SSD. 




\noindent\textbf{Real Control-Plane Deployment with an Emulated Large-Scale Cluster.}
For some experiments in \S\ref{sec:char} and \S\ref{sec:eval}, we use KWOK~\cite{kwok} v0.6.1
to simulate large numbers of worker nodes while retaining a real Knative-Kubernetes control plane, enabling high-scale evaluation of our serverless system without the prohibitive cost of provisioning physical clusters. KWOK supports modeling node and instance behavior to stress-test control-plane performance and experiment with scenarios, such as high-rate sandbox creation by the control plane~(\S\ref{sec:char_thru}) or configurable instance creation time~(\S\ref{sec:eval_sens_cluster_delays}), which is difficult to observe with real hardware outside a production environment. This approach uses real control-plane components with all their interactions, as in a large-scale production deployment, while modeling the load on worker nodes. This makes large-scale and configurable experiments~(\S\ref{sec:eval_large_scale}) feasible in a small-scale research setting.



\noindent\textbf{Software setup.} 
We use vHive~\cite{ustiugov:benchmarking}, i.e., Knative v1.13 running on top of Kubernetes v1.29. Regular Instances run atop containerd v1.6.18, while Emergency Instances execute in AWS Firecracker VMs v1.10.1, with snapshot-restore enabled. We limit each Knative instance's concurrency to 1 (i.e., maximum number of concurrent requests per instance), similarly to AWS Lambda~\cite{aws-lambda-scaling}. 
To prove the generality of our approach, we evaluate other values for per-instance concurrency separately in~\S\ref{sec:eval_cc}.
We assume all container images and VM snapshots are cached in memory on each node, similarly to prior work~\cite{cvetkovic:dirigent}.
To ensure measurement stability, we disable SMT and fix the CPU frequency to the base frequency,
and set the Load Balancer's (called Activator in Knative) replica count to 1, but ensure that it never becomes a bottleneck.\footnote{During the peak utilization periods in our experiments, Knative Activator uses less than one CPU core with a 99\textsuperscript{-th} percentile $<$10ms routing delay.}


\noindent\textbf{Workload.}
We use the In-Vitro~\cite{ustiugov:enabling} methodology
for representative trace sampling and load generation.
In all experiments, we use a 400-function trace sample with per-function IAT and duration distributions, chosen to apply the maximum possible load to the cluster without reaching 100\% CPU utilization at any point throughout the experiment. Only in \S\ref{sec:eval_large_scale}, we use a bigger trace with 2000 functions to evaluate the performance of a larger, 50-node cluster. We run experiments for an hour, discarding the first 20 minutes as a warm-up.
Similar to the prior work~\cite{cvetkovic:dirigent,ustiugov:enabling}, we use a synthetic spin-loop function with a programmed duration taken from the trace.
To confirm our findings with the synthetic functions, we then use a set of realistic serverless functions (encryption, authentication, fibonacci, and image-rotate) from vSwarm~\cite{schall:lukewarm}, which are written in Go, Python, and Node.js, in~\S\ref{sec:eval_real}. To achieve the load representative of a production setup, we replicate these functions so that each replica corresponds to one of the functions from the same 400-function sample derived from the Azure traces to fit the sample's execution duration and memory footprint distributions.


\noindent\textbf{Baselines.} 
We compare \sys to five state-of-the-art serverless systems. We use four baselines based on Knative~\cite{knative}, which is a Kubernetes-based serverless system widely used in commercial offerings~\cite{knative_offerings}. First is vanilla Knative (\textbf{Kn}) with an asynchronous control plane, which uses its default concurrency-based autoscaling policy with a 60-second autoscaling window.
We carefully configure the Knative deployment to maximize the baseline control plane's performance.\footnote{We increase concurrency and request rate limits to the Kubernetes API Server in Knative components and Kubernetes Controller Manager, and also increase the CPU and memory quotas for Knative core components. 
}
The second one is Knative-Synchronous (\textbf{Kn-Sync}), the Knative version we implement with synchronous instance creation similar to AWS Lambda. Specifically, we have modified Autoscaler to trigger new instance creations when it cannot find an available instance for a new request and retain the instances for a fixed period of inactivity. We use a 10-minute keep-alive period,
which is estimated to be the keep-alive duration for AWS Lambda~\cite{ustiugov:stellar}. 
We also use \textbf{Dirigent}~\cite{cvetkovic:dirigent}, a clean-slate serverless cluster manager with a high-performance asynchronous control plane, which is the state-of-the-art academic system.
Finally, (\textbf{Kn-NHITS}) and (\textbf{Kn-LR}) that use NHITS~\cite{challu:nhits} and lightweight linear regression prediction models (demonstrated as the most accurate predictors for FaaS by the prior work~\cite{joosen:how}), respectively, replacing the default Knative autoscaling policy. We train the models on the one-hour-long part of the trace that precedes the part used in the evaluation.


\noindent\textbf{Performance and Cost Metrics.}
We are using several metrics to evaluate the systems. Our main characteristics of the systems are function invocation performance and the costs that the serverless system incurs to provide that level of performance.
To evaluate performance, we use the geometric mean of the tail (99\textsuperscript{-th} percentiles) of per-function slowdown. Lower slowdown is better, with the slowdown of 1 meaning that the system's end-to-end response time is as low as measured in an unloaded system.\footnote{We first calculate per-invocation slowdown by dividing the end-to-end response time by expected execution duration. Then we compute the per-function 99\textsuperscript{-th} percentile of slowdown. Finally, we aggregate per-function slowdown with the geometric mean.} To estimate the cost-efficiency, we use the total memory footprint of all instances in the cluster, normalized to the total memory footprint of non-idle instances. We call this metric \emph{normalized cost}.
We also evaluate other cost sources, such as instance creation rates and the CPU cycles used by control-plane components.
We derive these metrics by collecting cluster-wide metrics with Prometheus~\cite{prometheus} and Kubernetes Metrics Server~\cite{k8s-metrics}. 

\section{Evaluation}
\label{sec:eval}


We evaluate \sys to answer the following questions: (1) How to set \sys' parameters for the best performance and lowest overhead? (\S\ref{sec:eval_sens}) (2) How does \sys's performance compare with existing systems? (\S\ref{sec:eval_perf}) (3) How does \sys's cluster resource utilization compare with existing systems? (\S\ref{sec:eval_resource_efficiency}) (4) How does \sys balance performance and cost trade-offs? (\S\ref{sec:eval_tradeoffs})

\subsection{\sys Sensitivity Studies}
\label{sec:eval_sens}

\begin{figure}
    \centering
    \subfigure[Keep-alive duration.]{\includegraphics[height=0.5\linewidth]{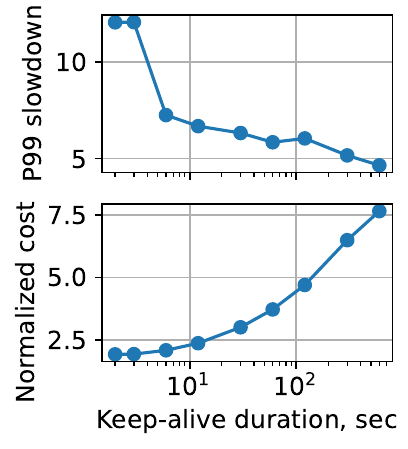}\label{fig:sensitivity:keepalive}} 
    \subfigure[Filtering percentile.]{\includegraphics[height=0.5\linewidth]{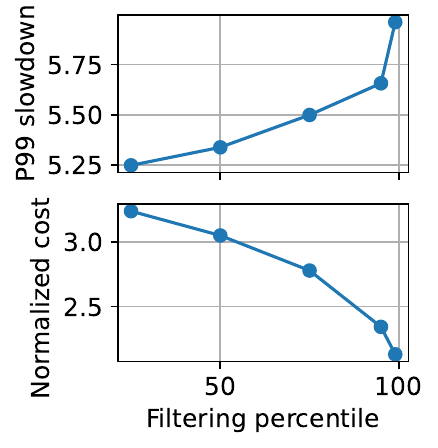}\label{fig:sensitivity:percentile}} 
    \caption{
    Effects of keep-alive duration (a) and filtering percentile (b) on \sys performance and cost.
    Results are collected under realistic samples from Azure Functions trace.
    }   
    \vspace{-10pt}
    \label{fig:sensitivity}
\end{figure}

As discussed in \S\ref{sec:des_standard}, \sys's performance is primarily affected by two parameters: (1) keep-alive duration and (2) metric filtering percentile.
Below, we study \sys sensitivity to these parameters and identify the sweet spot in the performance-cost trade-off to choose these parameters' values when running sampled production traces.

\noindent\textbf{Keep-Alive Duration.} 
The keep-alive duration determines how long a Regular Instance can remain idle before it is terminated. A smaller keep-alive increases the cold start probability, degrading performance. A higher keep-alive causes resource wastage, increasing cost. We sweep the keep-alive duration from 2s to 600s to measure its impact on system performance and cost. As Figure \ref{fig:sensitivity:keepalive} shows, when the keep-alive duration reaches 60s, further increasing it significantly increases the cost with minimal performance improvement. Thus, we set the keep-alive duration to 60s.


\noindent\textbf{Filtering Percentile.}
The filtering percentile determines the confidence threshold to create new function instances. Specifically, a lower filtering percentile increases the likelihood of creating new instances, leading to resource over-allocation and higher costs. A higher filtering percentile makes the system more conservative in creating new instances, increasing the number of cold starts and degrading performance. We vary the percentile from 25\% to 99\%. Figure \ref{fig:sensitivity:percentile} shows that \sys achieves the best balance between performance and cost when the filtering percentile is 50\%, which we use below. 

Given the obtained results, we choose the 60-second keep-alive period and the 50\textsuperscript{-th} percentile as metric filtering percentile in \sys as our parameters for evaluation.

\subsection{Control-Plane Performance Analysis}
\label{sec:eval_perf}


In this section, we evaluate \sys's control-plane performance from three perspectives: instance creation delays, scheduling delays, and sensitivity to instance creation delays.




\subsubsection{Instance Creation Delay Breakdown}

We first compare the creation delays of Regular and Emergency instances. \sys eliminates all interactions with the cluster manager and uses a pool of pre-allocated local IP addresses to create the Emergency Instance in under 150ms, nearly 10$\times$ faster than creating a Regular Instance in Knative (Figure~\ref{fig:k8s_inst_creation}).
Importantly, Regular Instance creation delays always occur off the critical path when handling invocations in \sys. 
Hence, \sys exposes only the cold-start delays of Emergency Instances to the application.



\subsubsection{Scheduling Delays Analysis}

\begin{figure}
    \centering
    \includegraphics[width=\linewidth]{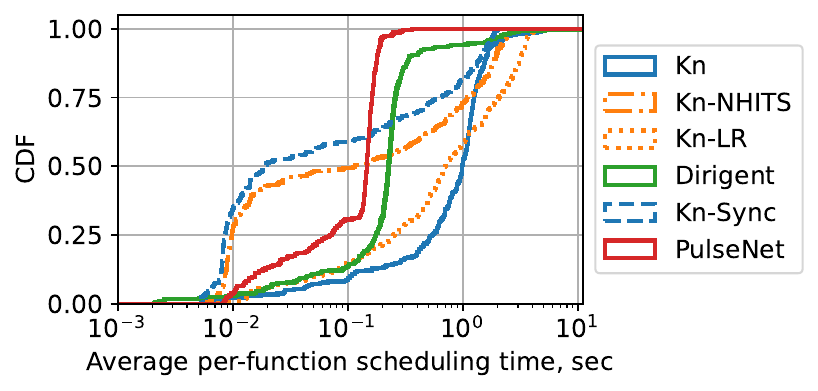}
    \caption{
    CDFs of average per-function scheduling delay in evaluated systems under a sampled production workload.
    }
    \label{fig:sched_delay}
\end{figure}

We continue the evaluation with the scheduling delays of different serverless systems.
Scheduling delay includes cold-start time, data-plane queuing, request routing, and load balancing. To measure the scheduling delay, we subtract the function's execution time from the end-to-end invocation latency. Figure~\ref{fig:sched_delay} shows the distribution of average scheduling delay for each system.

As shown in Figure \ref{fig:sched_delay}, Knative, Dirigent, and \sys show median delays of 1s, 200ms, and 150ms, respectively, matching their instance creation times. In addition to faster instance creation delays, \sys eliminates the worst-case latencies of up to 4 seconds that sporadically occur in Knative and Dirigent due to their window-based autoscaling policies.
Kn-NHITS and Kn-Sync, due to their high instance retention, have 40-50\% of functions that experience warm starts below 20ms. However, the remaining functions are delayed up to 2s due to slow instance creation. Kn-LR shows high latency, up to 4 seconds, indicating its high prediction error.
Compared with the above baselines, \sys reduces worst-case scheduling delays by quickly creating Emergency Instances when no Regular Instances are present or when all Regular Instances are busy, while minimizing cluster resource usage.

\subsubsection{Sensitivity to Various Instance Creation Delays}
\label{sec:eval_sens_cluster_delays}

\begin{figure}
    \centering
    \includegraphics[width=\linewidth]{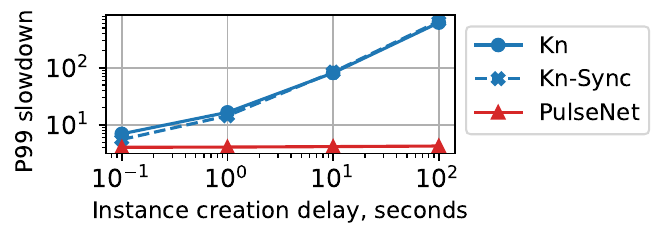}
    \caption{
    Slowdowns of the systems in cluster managers with different simulated instance creation delays, measured for sampled production workload. Lower is better.
    }
    \label{fig:sensitivity_cluster_manager}
\end{figure}


Different providers use different deployment and sandbox technologies, from regular VMs to FaaS-specialized solutions, to isolate Regular Instances. Hence, we evaluate the impact of their corresponding instance creation delays on the performance of the baseline systems and \sys.
We use KWOK~\cite{kwok} to set up the simulated instance creation delay from 100ms, as in container bootstrapping, to 100s, similar to the booting time of a general-purpose VM, and measure the performance of different serverless systems under the same sampled production trace.

Figure \ref{fig:sensitivity_cluster_manager} shows that increasing instance creation delay leads to a significant performance degradation for the baseline systems, whereas \sys shows no sensitivity because its fast-path control plane can quickly create Emergency Instances to handle burst traffic.

In summary, by creating Emergency Instances for excessive traffic, \sys effectively eliminates worst-case scheduling delays. Furthermore, based on this design, even in environments with high instance creation delays, \sys delivers more stable performance. 




\subsection{Control-Plane Resource Efficiency}
\label{sec:eval_resource_efficiency}


Next, we analyze \sys's control-plane resource efficiency from three perspectives: instance creation rate and CPU and memory utilization across the cluster.

\subsubsection{Instance Creation Rate}
\label{sec:eval_inst_creat}

\begin{figure}
    \centering
    \subfigure[Instance creat. rate.]
    {\includegraphics[width=0.31\linewidth]{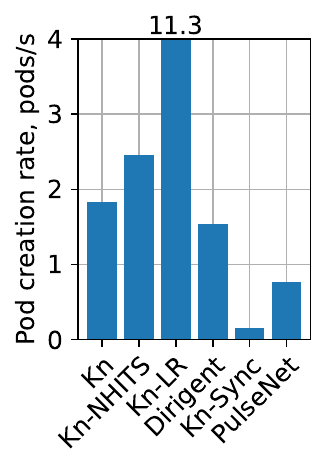}\label{fig:cold_starts}}
    \subfigure[CPU utilization breakdown.]
    {\includegraphics[width=0.68\linewidth]{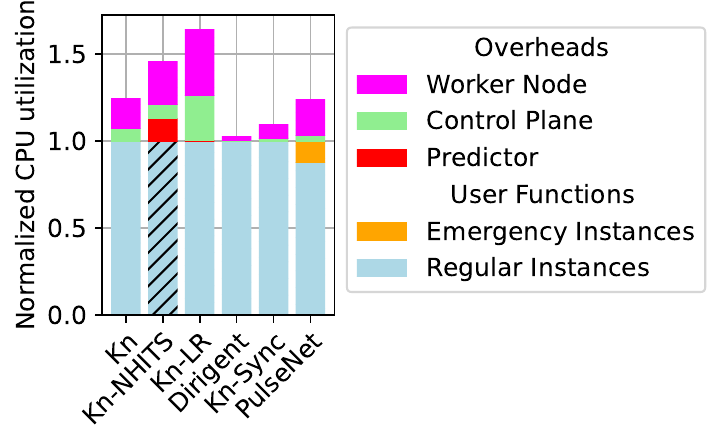}\label{fig:cpu_util}}
    \caption{
    Cluster-wide instance creation rates observed (a) and CPU utilization breakdown (b) in the systems under sampled production workload.
    }
    \vspace{-10pt}
    
\end{figure}

In this experiment, we measure the instance creation rate over time while executing the Azure Functions trace and show the results in Figure~\ref{fig:cold_starts}. A lower instance creation rate reduces the control-plane overhead and improves system stability.

As shown in Figure \ref{fig:cold_starts}, Kn-Sync has the lowest instance creation rate (0.1 instances/s) as it uses a long 10-minute keep-alive duration, allowing most instances to be reused. Knative and Dirigent show comparable instance creation rates (1.8 and 1.6 instances/s, respectively) as they employ similar autoscaling policies. The prediction model-based Kn-NHITS and Kn-LR have the highest instance creation rates (2.4 and 11.3 instances/s, respectively). This is because they need to adjust the number of instances based on prediction results frequently. Compared with Knative, \sys reduces the instance creation rate by 60\% to 0.8 instances/s. This is because \sys's filtering mechanism intelligently avoids creating unnecessary Regular Instances for functions with little, instead handling those with Emergency Instances. This design can reduce the computational overhead in the serverless system's control plane and improve system stability. We study the additional computational resources required to handle instance creation in the following section.






\subsubsection{Cluster-wide CPU Cycles Utilization Breakdown}
\label{sec:eval_cpu_util}



We define CPU overhead as additional CPU consumption beyond the resources required to serve user functions, which incurs costs for service providers. CPU overhead includes all control-plane components (sandbox management, traffic predictions, health and metric gathering), data-plane components (load balancer, ingress), and any computation overhead produced by prediction mechanisms.

As shown in Figure \ref{fig:cpu_util}, Emergency Instances account for only 10\% of total CPU usage by the instances. \sys has similar overhead as Knative (24\%) since it adds extra overhead for Emergency Instance management, while reducing the load on the control plane due to fewer instances created.
Prediction-based systems, Kn-NHITS and Kn-LR, have the highest CPU overhead due to additional compute resources required for predictions and handling more instance creations (we exclude their training time as an overhead): 45\% and 64\%, respectively.
Dirigent can handle the same instance creation rate as Knative with 2\% computational overhead due to its lighter cluster management implementation. Kn-Sync, due to its longer keep-alive period, reduces the frequency of instance creation, thereby decreasing CPU overhead to 9\%. Although \sys CPU overhead is higher than that of Kn-Sync, \sys achieves a better balance of resource usage by trading slightly increased CPU overhead for significantly reduced memory usage. We will analyze this in depth in \S\ref{sec:eval_memory_util}. 

\subsubsection{Memory Usage}
\label{sec:eval_memory_util}

\begin{figure}
    \centering
    \includegraphics[width=0.85\linewidth]{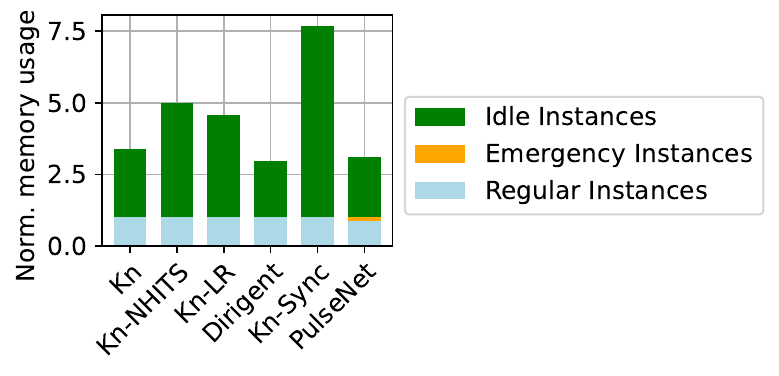}
    \caption{
    Memory usage for different serverless systems under sampled production workload normalized to the number of active instances. Lower is better.
    }
    \label{fig:mem_usage}
    \vspace{-10pt}
\end{figure}


In this section, we evaluate the memory usage of various serverless systems by replaying the Azure Functions trace. 
We show our results in Figure \ref{fig:mem_usage}.
Kn-Sync exhibits the highest memory usage because it employs a long keep-alive period of 10 minutes to handle requests, using 7$\times$ more memory for idle instances than for active ones.
The prediction accuracy of Kn-NHITS and Kn-LR leads to overprovisioning of instances of 5$\times$ and 4.5$\times$, respectively.
Dirigent, due to its fast instance creation speed, can create instances more quickly, reducing queuing and preventing subsequent overreaction by the autoscaler.
\sys uses our proposed metric-filtering technique, which allows some requests to be handled by an Emergency Instance without creating a long-lived Regular Instance. These Emergency Instances account for 10\% of the total non-idle instance memory usage. Compared with Knative and Kn-Sync, \sys improves memory utilization by 8\% and 60\%, respectively.
\sys uses 5\% more memory for instances than Dirigent, which could be improved with better metric filtering mechanisms~(\S\ref{sec:des_standard}).






In summary, with our proposed dual-path control-plane architecture and intelligent filtering mechanism, \sys achieves the best balance between memory and CPU resources among Kubernetes-compatible systems.
Specifically, \sys reduces the memory utilization by 8-60\% compared to the baselines (and only 5\% more than the FaaS-specialized Dirigent) and imposes negligible CPU overhead. 
\subsection{Performance \& Cost Trade-off Analysis}
\label{sec:eval_tradeoffs}


In this section, we compare the performance and resource usage of different serverless systems. We also validate these results in a large-scale cluster with emulated worker nodes.


\begin{figure}
    \centering
    \includegraphics[width=0.9\linewidth]{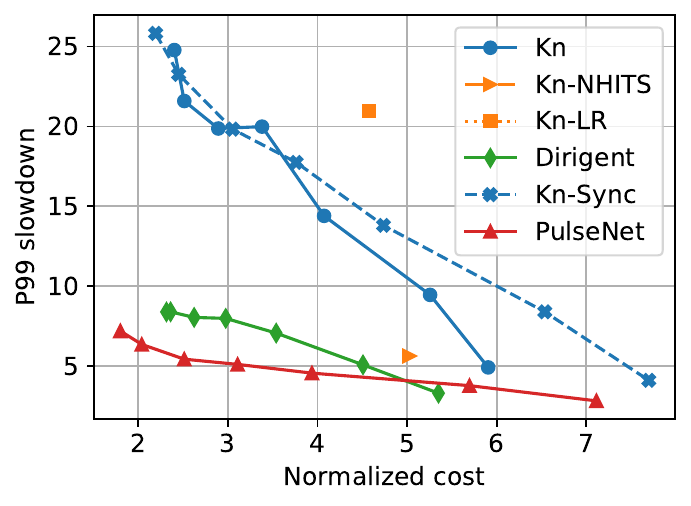}
    \caption{
    Performance-cost trade-off comparison for different serverless systems under sampled production workload. Lower is better for both axes.
    }
    \label{fig:perf_v_cost}
    \vspace{-10pt}
\end{figure}

\noindent\textbf{Experiments with a Real System.}
\label{sec:eval_real_system}
%
In this section, we evaluate the trade-off between performance and cost for different systems. 
We obtain the performance-cost trade-off by varying instance retention-related parameters (\textit{e.g.,} keep-alive period, autoscaling window) from 6 seconds to 10 minutes.
As shown in Figure \ref{fig:perf_v_cost}, \sys achieves the best tradeoff between performance and cost. Compared with Knative and its variants (such as Kn-NHITS and Kn-LR), \sys achieves higher performance at lower resource costs.
Specifically, \sys outperforms Kubernetes-compatible systems with synchronous control planes by 1.5-3.5$\times$ at 8-70\% lower cost, and surpasses asynchronous counterparts by 1.7-3.5$\times$ at 3-65\% lower cost.
\sys achieves 35\% faster end-to-end performance at a comparable cost to the Dirigent system.
The reason behind \sys's optimal tradeoff is its dual-track control-plane design. This design effectively manages sustainable traffic with Regular Instances and excessive traffic with Emergency Instances, thereby achieving higher performance with fewer resource costs.






\noindent\textbf{Large-Scale Experiments.}
\label{sec:eval_large_scale}
Then, we use the same methodology and metrics 
to evaluate the performance-cost trade-offs in the large-scale cluster by simulating 50 worker nodes with KWOK and running real control-plane components.
The experimental results show that at this larger scale, \sys still demonstrates significant improvements compared with state-of-the-art systems. 
Specifically, \sys outperforms Kubernetes-compatible systems with both synchronous and asynchronous control planes by up to 3$\times$ at up to 3$\times$ lower cost.
As the cluster size increases, traditional systems experience severe control-plane congestion. Additionally, system configurations that performed well in smaller clusters can overload the control plane under larger workloads. \sys effectively addresses this problem through the filtering mechanism between its expedited and standard tracks.

\subsection{Generalizability of \sys}

\noindent\textbf{Sensitivity to Per-Instance Concurrency.}
\label{sec:eval_cc}
%
%
As some providers, such as Google Cloud Run~\cite{gcr}, recommend setting up per-instance concurrency~(CC) larger than 1, we compare \sys performance and cost to the Knative baseline with larger CC.
For the Knative baseline, increasing CC from 1 to 10 improves performance by about 20\%, with the speedup saturating at CC of 5. Hence, a higher CC value can modestly shrink the gap in performance between \sys and the baseline (as CC does not affect the expedited track of \sys, \sys performance does not change), but the latter remains more than 3$\times$ faster. Furthermore, operating Knative at CC of 5 increases the overall operational cost by 3$\times$, assuming that concurrency-enabled instances incur proportionally higher resource costs. 



\noindent\textbf{Real-World Applications.}
\label{sec:eval_real}
\begin{figure}
    \centering
    \includegraphics[width=\linewidth]{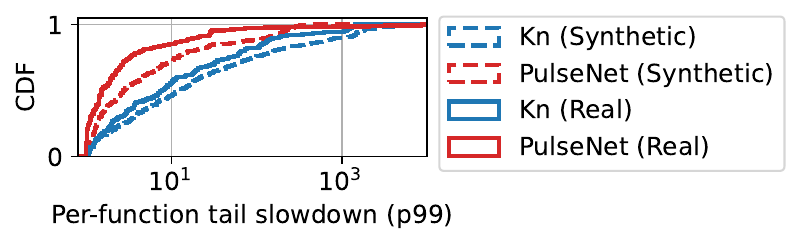}
    \caption{
    Tail slowdown for real-world functions and synthetic applications under production workload.
    }
    \vspace{-10pt}
    \label{fig:real_slowdown}
\end{figure}
We verify the generality of our findings by running real-world workloads from vSwarm written in Go, Python, and Node.js.
We observe a 4.3$\times$ reduction in tail slowdowns, as shown in Figure~\ref{fig:real_slowdown}, with a negligible impact on the number of instances in \sys relative to the Knative baseline. 
This performance gain is similar to previous experiments, however, with a larger performance improvements over the baseline with real functions compared to with synthetic ones, confirming the effectiveness of \sys in real-world deployments.\footnote{
We attribute those differences to the limitations of mapping 400 functions from Azure Trace to a set of replicated 11 vSwarm functions.
}

\section{Related Work}









\textbf{Cluster manager designs.} 
Both practitioners and academics explore many designs for datacenter cluster managers primarily targeting long-running services and microservices~\cite{k8s,hindman:mesos,vavilapalli:yarn,schwarzkopf:omega,isard:quincy,gog:firmament}. Some works~\cite{ousterhout:sparrow,schwarzkopf:omega} explore trade-offs between centralized and decentralized cluster manager designs. Other works~\cite{delimitrou:paragon,delimitrou:tarcil,delimitrou:quasar} study the interference among the services co-located in the same cluster.
The advent of Borg~\cite{verma:borg,tirmazi:borg} has mostly closed the research area for general-purpose cluster managers, making its open-source alternative Kubernetes~\cite{k8s} the system of choice for long-running services and the foundational ecosystem for cloud offerings. 
\sys is fully compatible with the cluster managers, such as Kubernetes, benefiting from their mature ecosystem and optimizations.
Although we show that conventional control planes based on Kubernetes deliver enough throughput to avoid queuing delays when creating instances at the datacenter scale, KOLE~\cite{zhang:kole} adapts Kubernetes for large-scale edge deployments but foregoes run-time instance creation essential for highly-dynamic serverless workloads.

\textbf{FaaS-Oriented Control-Plane Optimizations.} 
Recent works explore FaaS-specialized control planes, specifically focusing on FaaS workloads and often excluding the BaaS services from their consideration. Some works~\cite{shahrad:serverless,liu:jiagu,roy:icebreaker,joosen:how,mittal:mu} focus on predicting the arrival time or request concurrency of future invocations to pre-allocate function instances ahead of time. Il\'uvatar~\cite{fuerst:iluvatar} reduces the scheduling overhead for warm starts. 
\sys is complementary to the above works, as it supports predictor-informed scaling in the standard track, and uses the conventional components in its warm-start path. 
Finally, other works~\cite{cvetkovic:dirigent,singhvi:atoll,mvondo:ofc,fuerst:faascache} take a radical approach designing FaaS-only scalable high-throughput control planes, foregoing the benefits of FaaS and BaaS application components colocation in the same cluster.

\textbf{BaaS-Oriented Control-Plane Optimizations.}
Several works have proposed BaaS-oriented control-plane designs tailored for serverless applications.
Some works~\cite{romero:faast,klimovic:pocket,mahgoub:sonic} optimize BaaS control planes for ephemeral storage by automatically scaling cache resources and dynamically selecting storage backends based on application workload patterns. 
Other works~\cite{Jesalpura:Shattering,yu:following,du:molecule} propose fast technologies to facilitate data movement across functions, transparently to the application.
\sys is complementary to these BaaS-specialized control planes, which can potentially be implemented as Kubernetes scheduler plug-ins.
Furthermore, SPRIGHT~\cite{qi:spright} and Nightcore~\cite{jia:nightcore} mitigate overheads in data movement between co-located instances by replacing TCP/IP networking with eBPF and IPC, respectively. FUYAO~\cite{liu:fuyao} extends this approach to optimize intra-node communication by eliminating the TCP/IP stack, in addition to DPU-accelerated inter-node data movement. Similarly, Apiary~\cite{kraft:apiary} places the function execution into the database's user-defined functions. These works underscore the benefits of data locality, which can be achieved through scheduling policies in conventional cluster managers such as Kubernetes~\cite{bazzaz:preventing,zhang:faster}.

\textbf{Other optimizations for serverless systems: cold-start delays, microvms, and hypervisors.}
Many recent works~\cite{szekely:sigmaos,ruan:quicksand,kuchler:dandelion,segarra:granny,fried:junction,huang:trenv} redesign the runtimes to improve the performance and efficiency of serverless and cloud deployments. Other works~\cite{fu:serverlessllm,zeng:medusa} apply similar approaches for GPU-centric workloads. These works are orthogonal to \sys and can be seamlessly incorporated into the \sys tracks.

\section{Conclusion}

This work resolves the long-standing serverless trade-off between the performance of clean-slate systems and the compatibility of conventional managers. We introduce \sys, a hybrid architecture built on the insight that FaaS traffic is bimodal—split between sustainable and excessive patterns. Its dual-track control plane melds a standard, compatible path for sustainable traffic with an expedited path using disposable fast instances for excessive bursts. \sys proves that serverless platforms can achieve high performance and full compatibility simultaneously, without compromise.

\bibliographystyle{plain}
\bibliography{./bibcloud/gen-abbrev,ref,dblp}


\end{document}